\documentclass{PoS}

\title{Details of a staggered fermion data analysis}

\ShortTitle{Details of a staggered fermion data analysis}

\author{\speaker{Maximilian Ammer}\\
         University of Wuppertal, D-42119 Wuppertal, Germany\\
        E-mail: \email{\,ammer@uni-wuppertal.de}}

\author{Stephan D\"urr\\
        University of Wuppertal, D-42119 Wuppertal, Germany \\
        IAS/JSC Forschungszentrum J\"ulich, D-52425 J\"ulich, Germany \\
        E-mail: \email{\,durr@itp.unibe.ch}}

\abstract{We present technical details of an analysis of pseudo-scalar data from a
QCD simulation with staggered fermions. The data were obtained close to
the physical point with an inverse lattice spacing of about 3 GeV, and
$N_f=2+1+1$. We compare different methods of extracting effective masses
and decay constants in lattice units. The results of several correlated
and uncorrelated fitting methods are compared, both on the simulated data
set, and on a synthetically generated data set.}

\FullConference{37th International Symposium on Lattice Field Theory - Lattice2019\\
		16-22 June 2019\\
		Wuhan, China}

\usepackage{amsmath}
\newcommand{\meff}{M_\mathrm{eff}}
\newcommand{\feff}{F_\mathrm{eff}}
\newcommand{\tmin}{t_\mathrm{min}}
\newcommand{\tmax}{t_\mathrm{max}}

\begin{document}

\section{Introduction}
The first step in extracting hadron masses from lattice data is to determine the effective masses and decay constants in lattice units (i.e.\ before any physical scale setting) along with their statistical and systematic errors. To illustrate this for the case of pseudo-scalar mesons we use one ensemble generated by the BMW-collaboration \cite{Borsanyi:2016lpl,Borsanyi:2013bia} with $N_f=2+1+1$ quark flavours in the staggered fermion representation with $\beta=4.0126$ (or lattice spacing $a\approx 0.063$ fm) and masses close to the physical point. The six pseudo-scalar channels will be denoted by their valance quark content as \texttt{ll,ss,cc,ls,lc} and \texttt{sc}.
The raw correlator data come in $N\times T= 441\times 144$ matrices, where $N$ is the number of configurations and $T$ the temporal extent of the lattice. Because of the periodic boundary conditions of the lattice, the data can be symmetrized (discarding the value at $t=0$) to reduce noise. Therefore the central value correlator $C(t)$ that will be used in  the analysis is given by the mean over all configurations and is defined on $t\in \{1,\dots ,T/2=72\}$.
\begin{figure}[b]
\center
\includegraphics[scale=0.45]{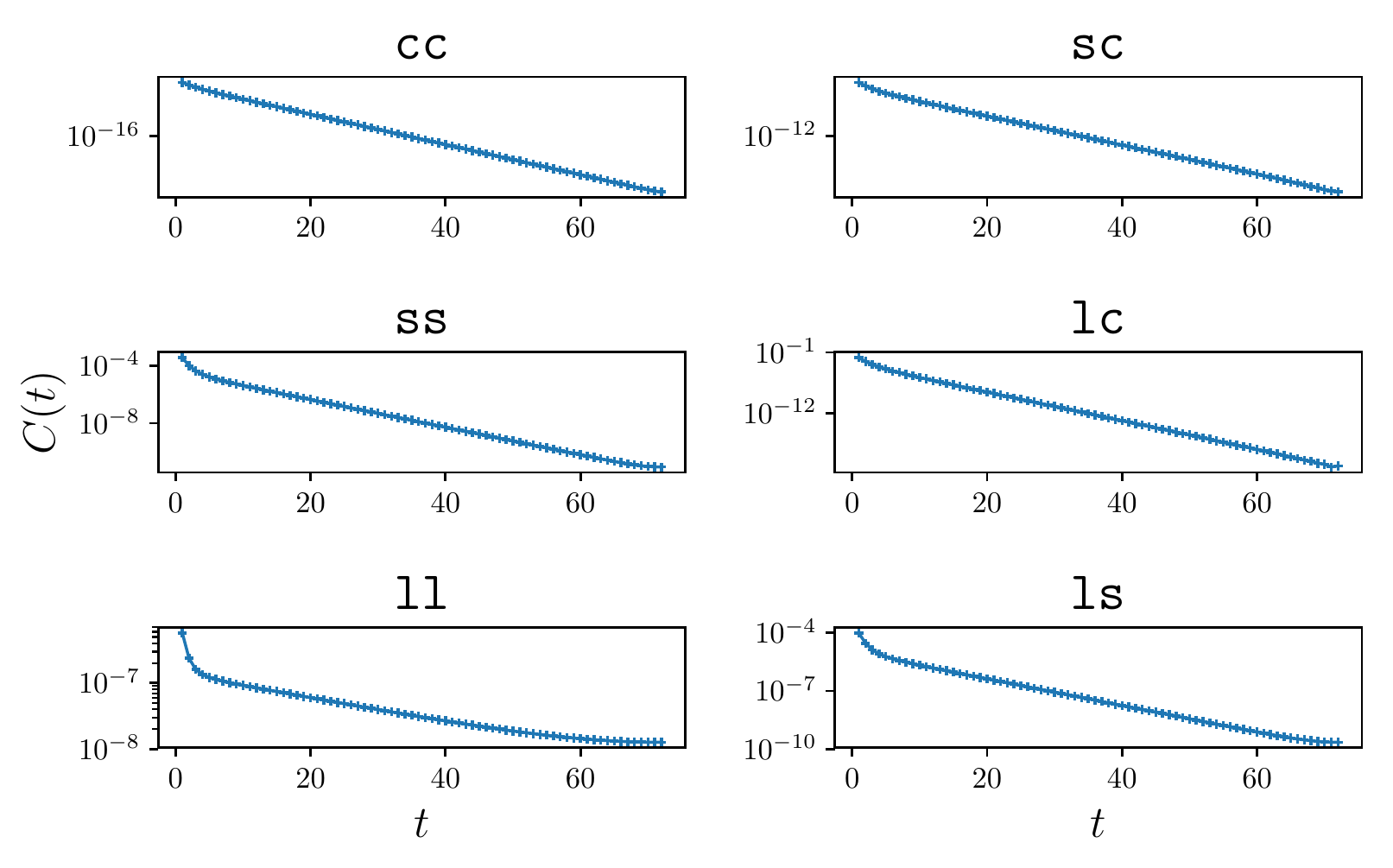}
\includegraphics[scale=0.45]{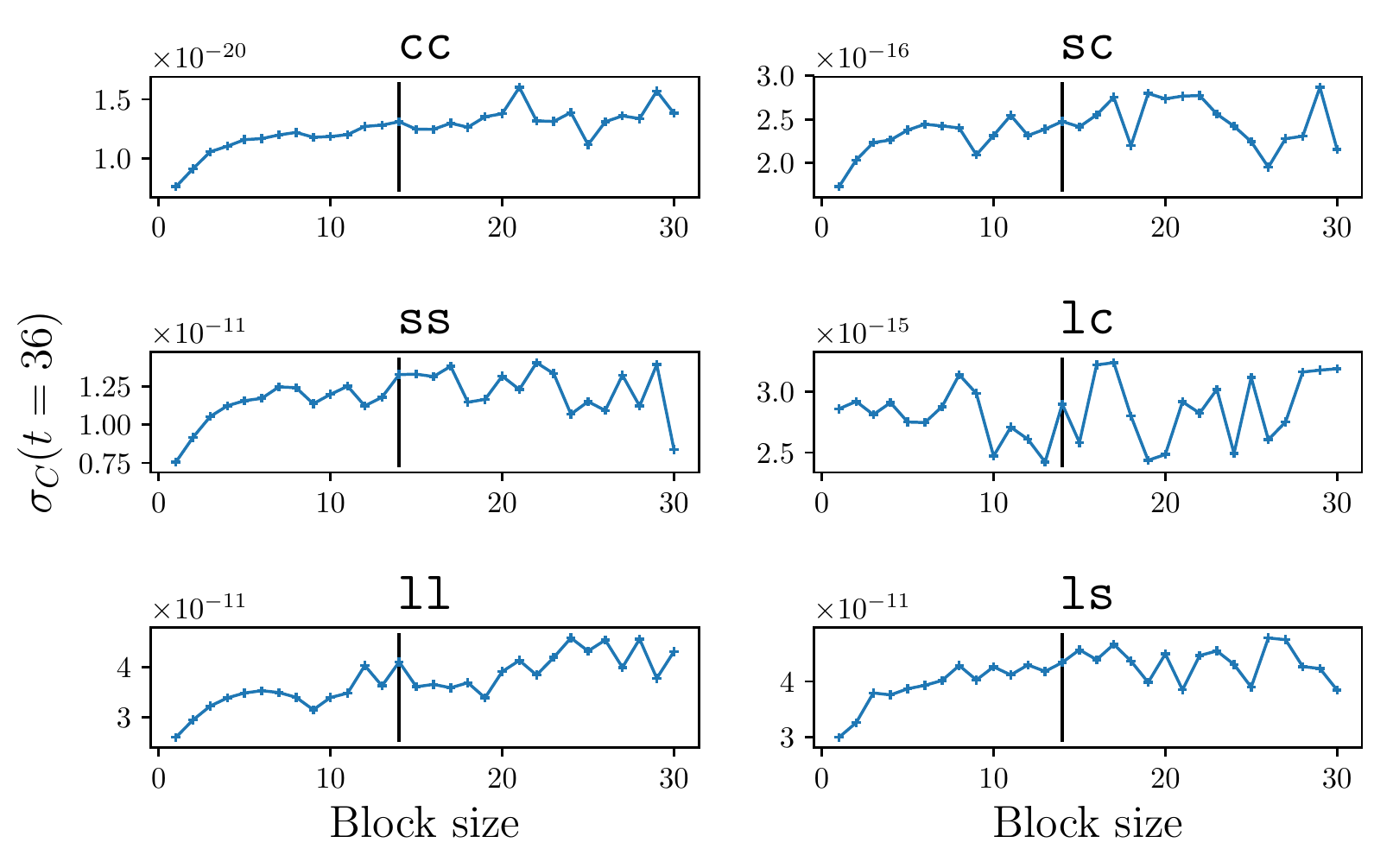}
\caption{Correlator data after symmetrization about $T/2=72$ for the six pseudo-scalar channels, plotted on a logarithmic scale (left).  Jackknife errors $\sigma_C$ at fixed $t=36$ as a function of the block size (right). A block size of 14  (indicated by the vertical lines) was chosen for the analysis. \label{fig1} }
\end{figure}

To calculate statistical errors and covariances we use the jackknife re-sampling technique. First, to minimize the auto-correlation between subsequent configurations, the data is blocked into $n$ blocks of length $\ell_\mathrm{bin}$ such that $N=n\cdot\ell_\mathrm{bin}+n_\mathrm{disc}$ and the first $n_\mathrm{disc}$ configurations are discarded. Each block is averaged to obtain the blocked values $C_i^\mathrm{bl}(t), i\in\{1,\dots,n\}$, then the jackknife samples are calculated according to 
\begin{align*}
C_i^\mathrm{jk}(t)=\frac{1}{n-1}\sum\limits_{j\neq i}^n C_j^\mathrm{bl}(t).
\end{align*}
With them the squared jackknife errors are given by 
\begin{align*}
\sigma_{C(t)}^2=\frac{n-1}{n}\sum\limits_{i=1}^n\left( C_i^\mathrm{jk}(t)-\left(\frac{1}{n}\sum\limits_{j=1}^n C_j^\mathrm{jk}(t)\right) \right)^2\cdot \left[\frac{N-n_\mathrm{disc}}{N}\right] .
\end{align*}
The last term in square brackets corrects for the discarded configurations.
In Figure \ref{fig1} the Jackknife errors are plotted as functions of the block size $\ell_\mathrm{bin}$. When they reach a plateau the block size is large enough for the blocks to be considered independent. In our case $\ell_\mathrm{bin}=14$ was chosen.

\section{Local definitions of $\meff$ an $\feff$}
\begin{figure}[b]
\includegraphics[scale=0.5]{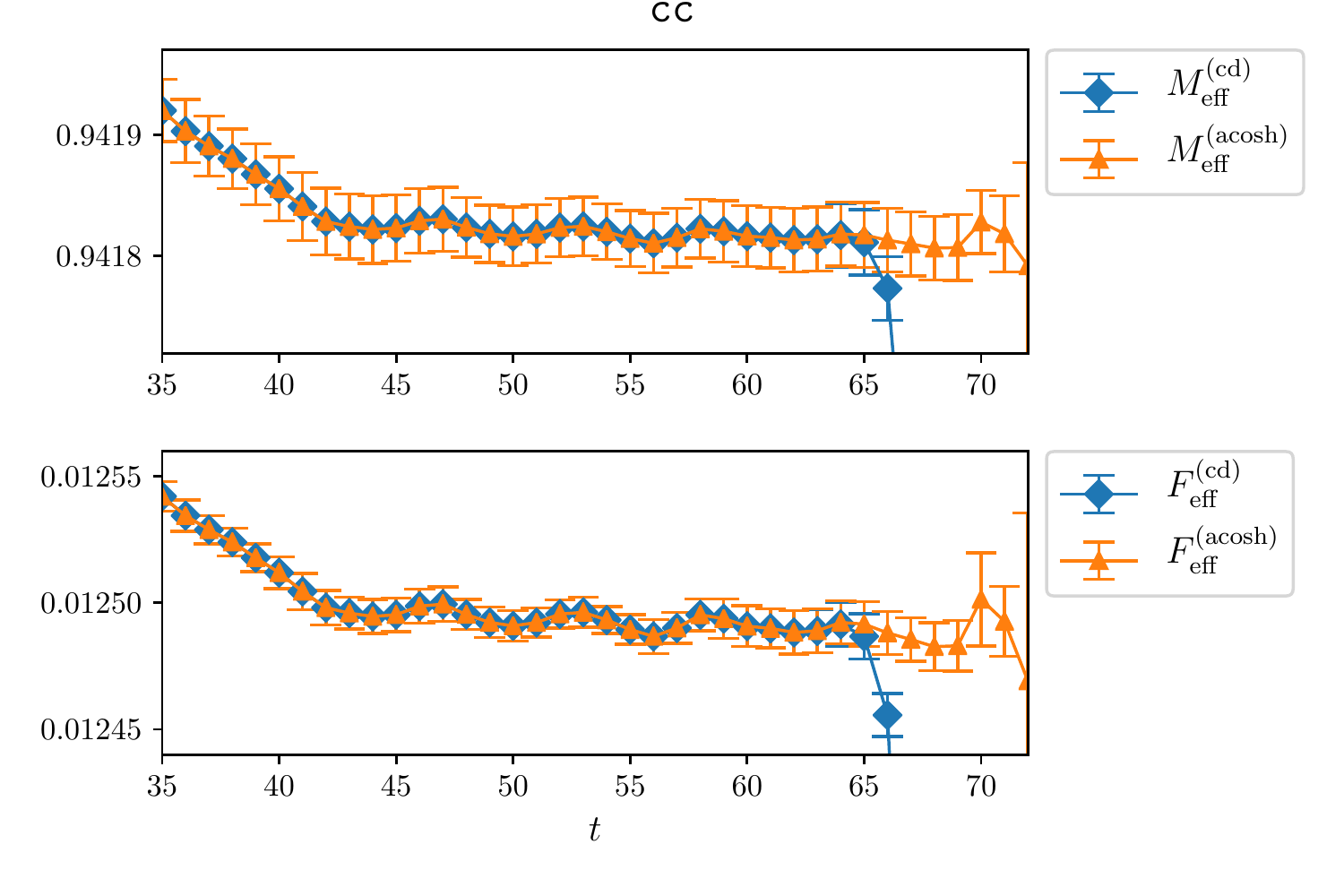}
\includegraphics[scale=0.5]{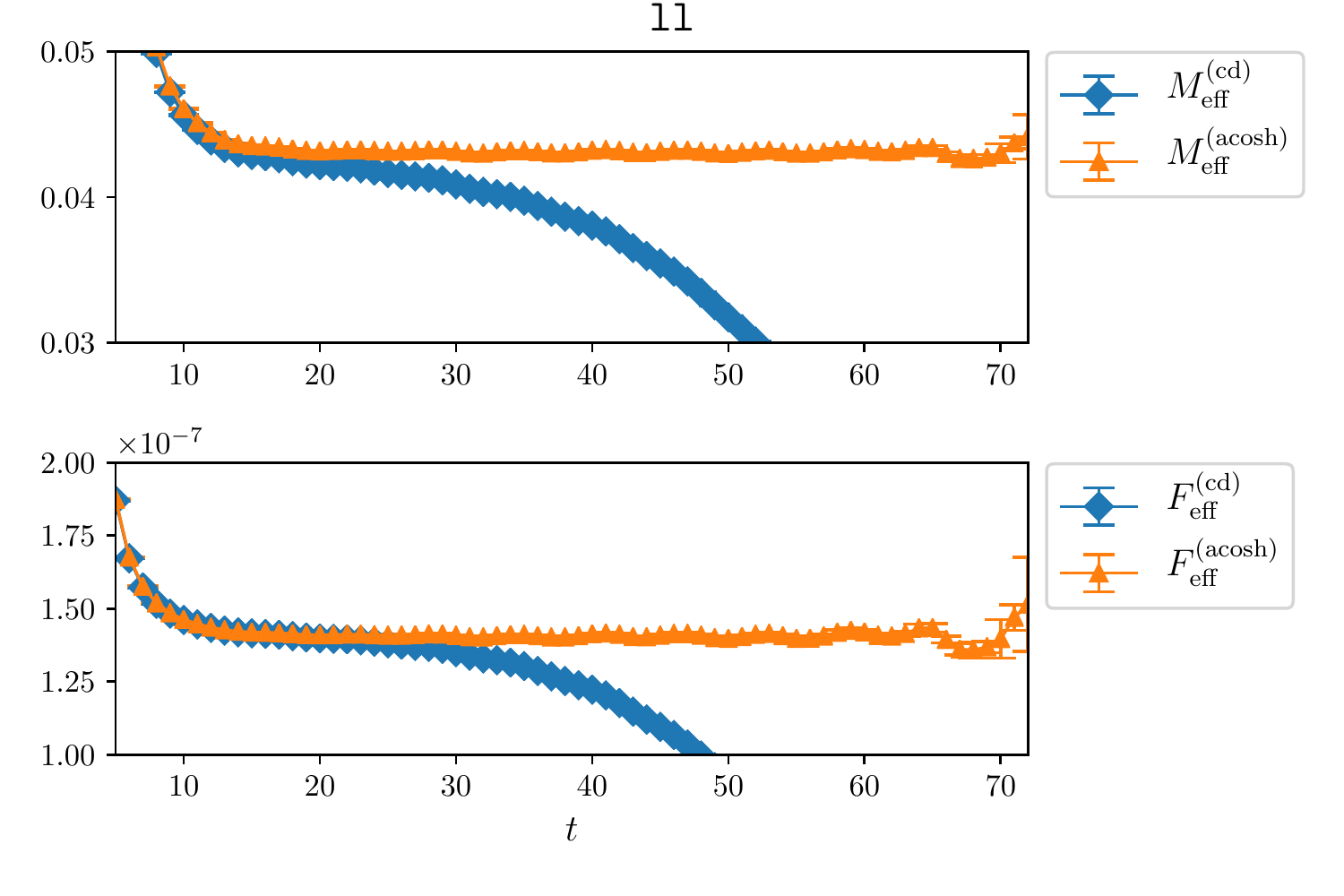}
\includegraphics[scale=0.5]{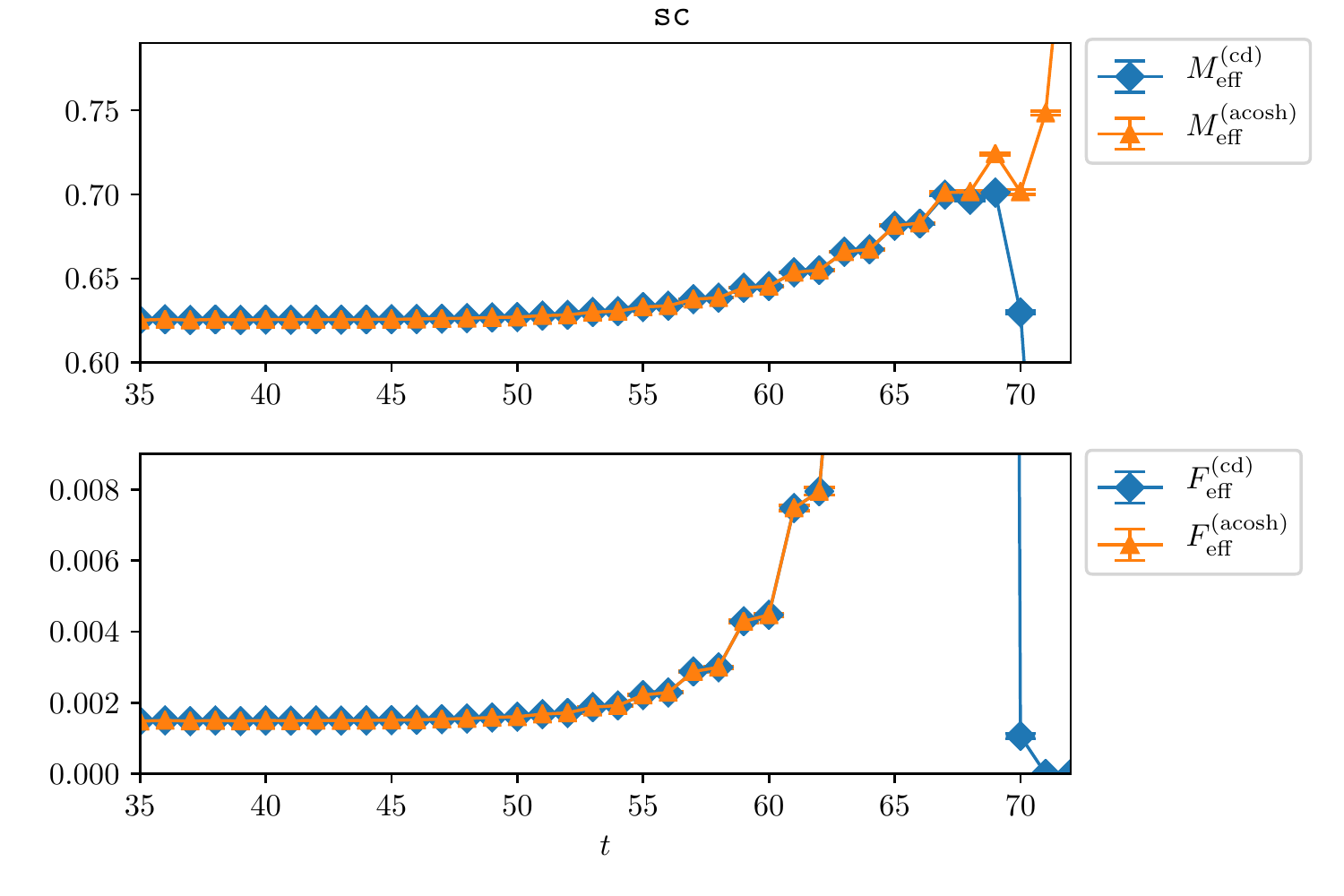}
\includegraphics[scale=0.5]{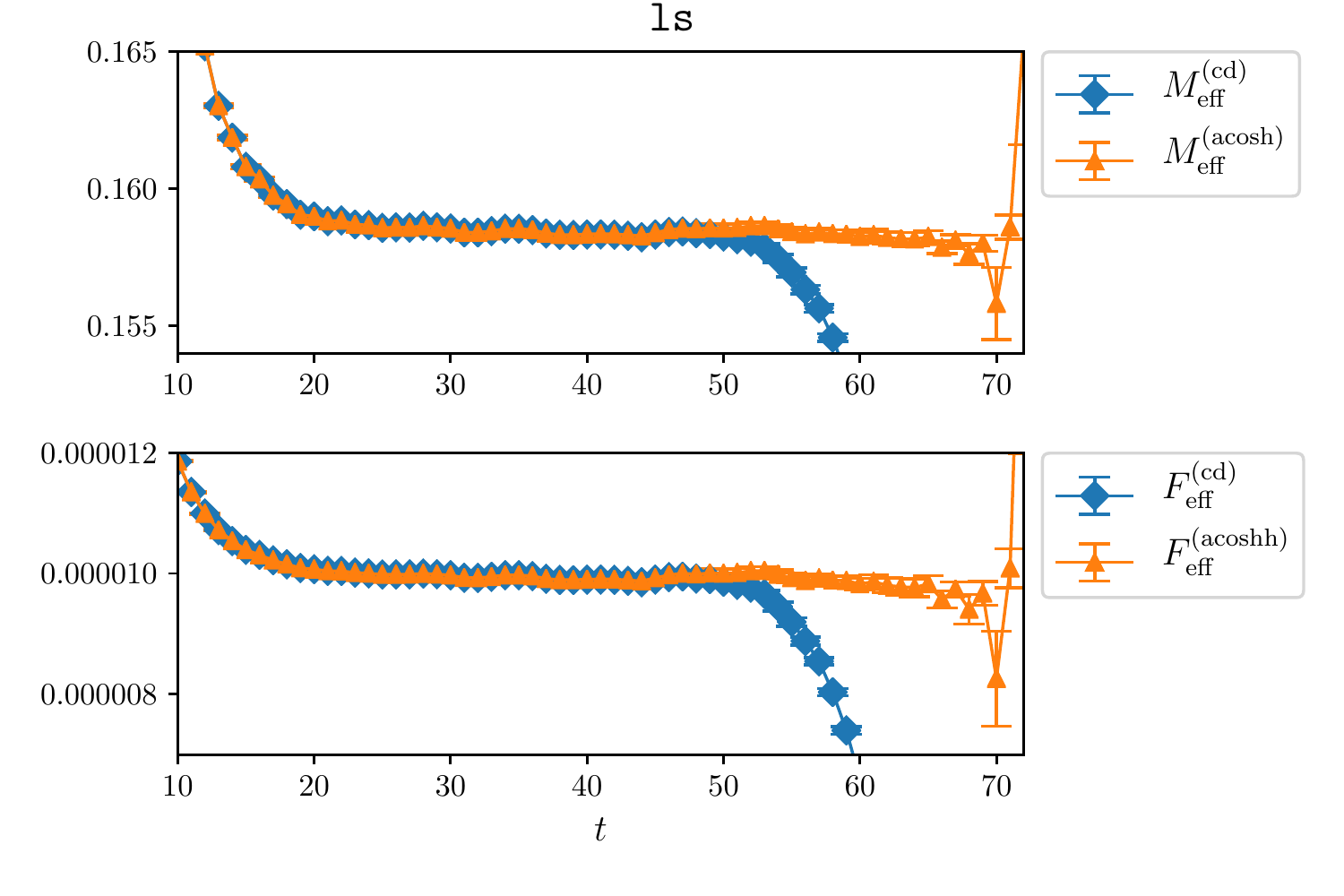}
\caption{Comparison of different definitions of the effective mass $\meff$ and effective decay constant $\feff$ for the \texttt{cc}-channel (top left), \texttt{ll}-channel (top right), \texttt{sc}-channel (bottom left) and \texttt{ls}-channel (bottom right).\label{fig2}}
\end{figure}
The effective mass $\meff$ and effective decay constant $\feff$ are extracted by fitting the correlator $C(t)$ itself or by fitting local definitions of $\meff$ and $\feff$. Because we are analysing staggered fermion data, the channels with a difference in quark masses (\texttt{sc,lc,ls}) exhibit an oscillatory behaviour with period $2 a$ (staggered fermion oscillations). Therefore we are specifically interested in a local $\meff$, which is defined on two time-slices with distance $2 a$ as this will smooth out these oscillations considerably.
We start by considering only the ground state and ignoring the periodic boundary conditions. Thus we assume (for now) the following form for the correlator:
\begin{align}\label{eq1}
C(t)=\feff\cdot e^{-\meff\cdot t}
\end{align}
A suitable local value for $\meff$ would then be the central derivative (cd) definition
\begin{align}
\meff^\mathrm{(cd)}(t)=\frac{\ln C(t-a)-\ln C(t+a)}{2a}.
\end{align}
More realistically the correlator has the form of a cosh function
\begin{align}\label{eq2}
C(t)=\feff \left(e^{-\meff \cdot t}+e^{-\meff\cdot (T-t)}\right)
\end{align}
\\
due to the backwards propagating contribution
and a nice definition for $\meff(t)$ is \cite{Toth-pc}:
\begin{align}
\meff^\mathrm{(acosh)}(t)=\frac {1}{2a} \left(\cosh^{-1}\left(\frac{C(t-1)}{C(T/2)}\right)-\cosh^{-1}\left(\frac{C(t+1)}{C(T/2)}\right)\right).
\end{align}
The corresponding $\feff^\mathrm{(cd)}(t)$ and $\feff^\mathrm{(acosh)}(t)$ are calculated by inverting equations (\ref{eq1}) and (\ref{eq2}) respectively.
In Figure \ref{fig2} the two local definitions are compared. Especially in the channels involving a light quark $\meff^\mathrm{(cd)}$ falls off for $t\to T/2$ and in the case of the pion channel (\texttt{ll}) it doesn't even form a plateau before doing so. The acosh-definition, on the other hand, seems to be doing a perfect job. In the $\texttt{sc}$-channel, finally, we see the signature of insufficient solver precision for $t\to T/2$. 
\section{Fits of $\meff(t)$,  $\feff(t)$ and $C(t)$}
\begin{figure}
\center
\includegraphics[scale=0.545]{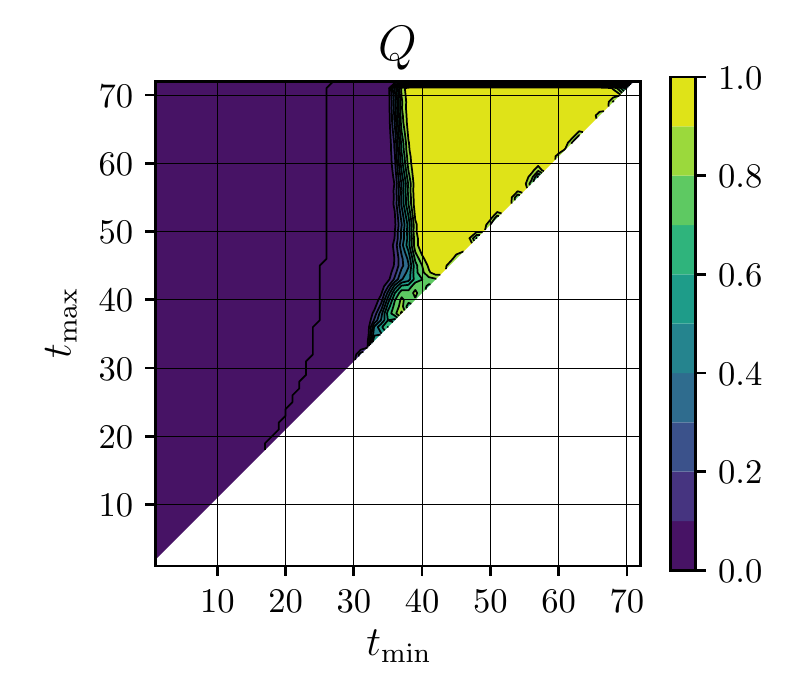}
\includegraphics[scale=0.545]{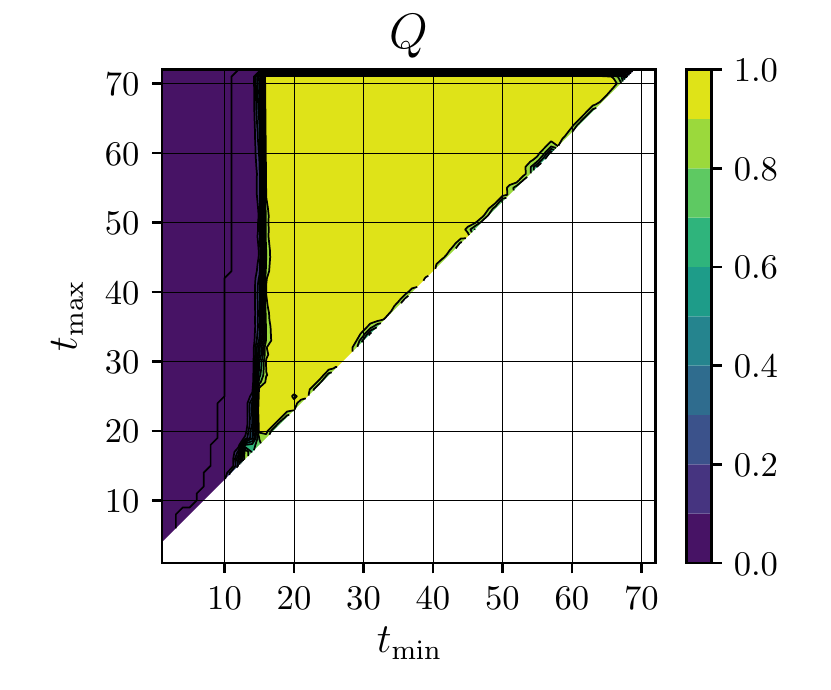}
\caption{The $Q$-value as a function of the fit interval $[\tmin,\tmax]$ for fits of $\meff(t)$ in the \texttt{cc}-channel. Constant fit function $f(t)=c_0$ (left) and a fit function including contributions from excited states $f(t)=c_0+c_1 e^{-c_2 t}$ (right). The $Q$-values are based on the uncorrelated $\chi^2$.\label{fig3}}
\end{figure}

\begin{figure}
\includegraphics[scale=0.45]{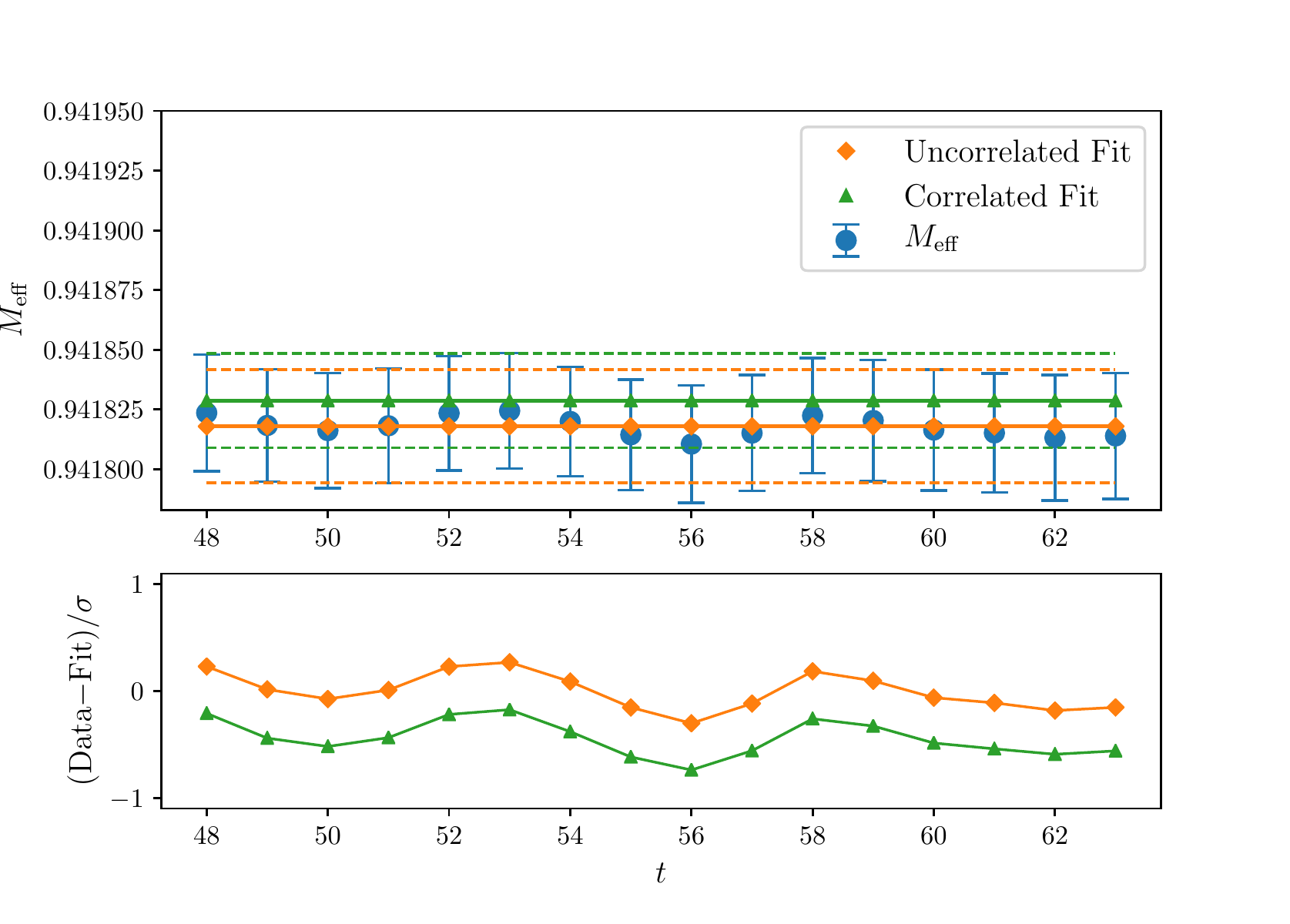}
\includegraphics[scale=0.45]{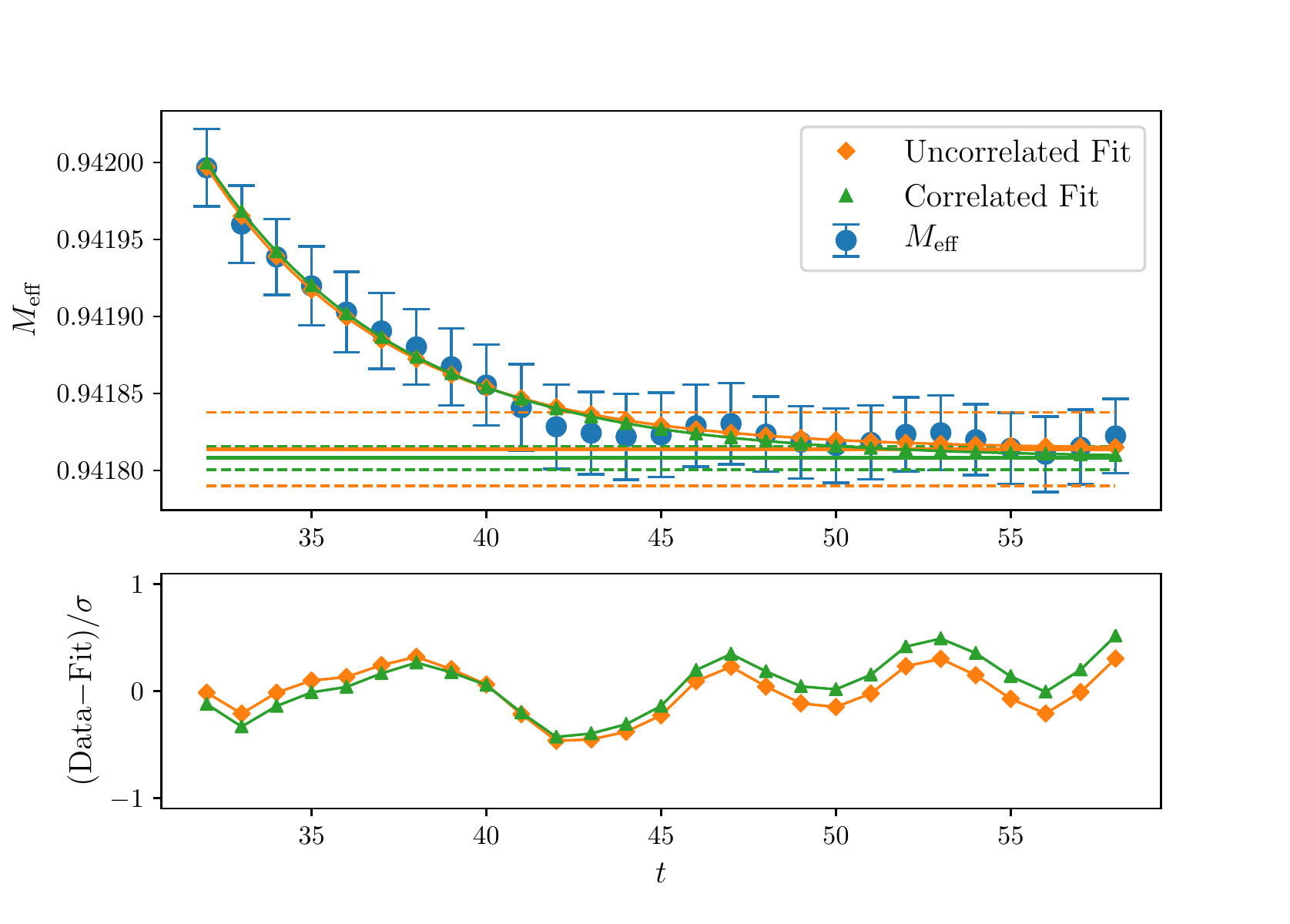}
\caption{Uncorrelated and correlated fits of $\meff(t)$ in the \texttt{cc}-channel, using
\texttt{const}: $f(t)=c_0$ (left) and  \texttt{constexp} : $f(t)=c_0+c_1 e^{-c_2 t}$ (right).  The final values of $c_0$ (with error) and the local bias are also indicated.\label{fig4}}
\vspace*{\floatsep}
\includegraphics[scale=0.45]{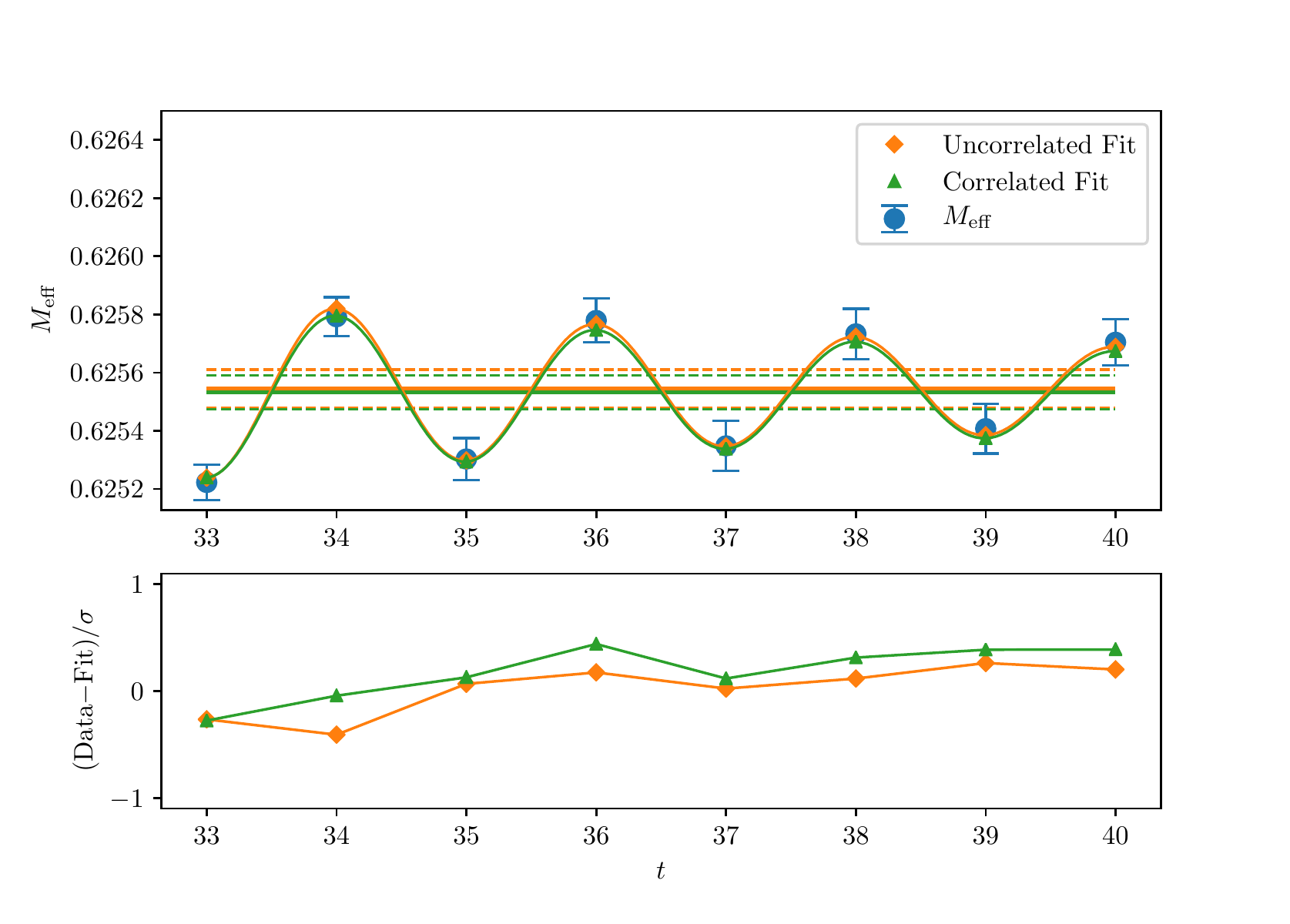}
\includegraphics[scale=0.45]{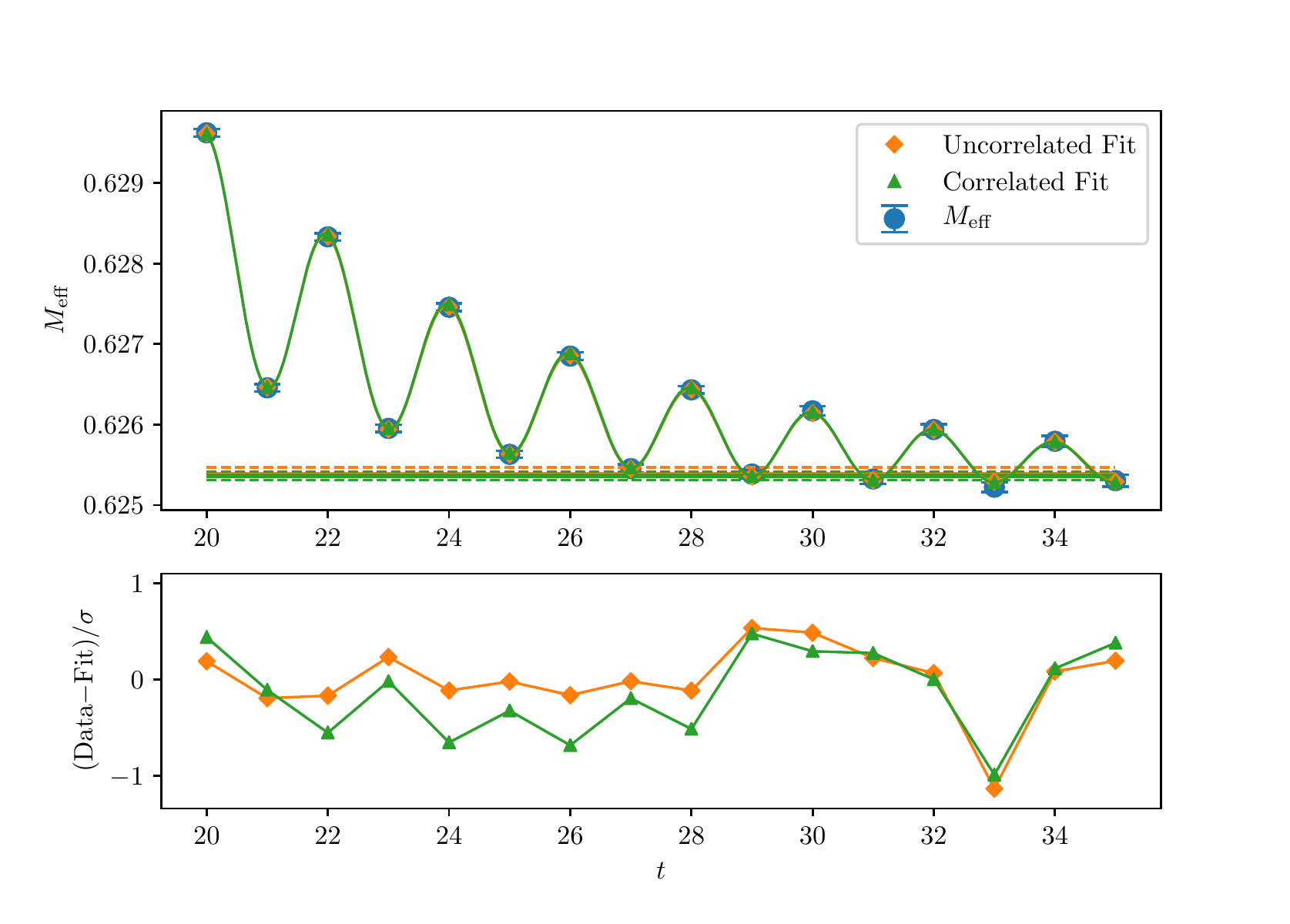}
\caption{Same as Figure \ref{fig4} but in the \texttt{sc}-channel and thus with an oscillatory term. Hence the fit finctions are \texttt{constcos}: $f(t)=c_0+c_1 e^{-c_2 t}\cos(\pi t)$ (left) and \texttt{constexpcos}:   $f(t)=c_0+c_1 e^{-c_2 t}+c_3 e^{-c_4 t}\cos(\pi t)$ (right).\label{fig5}}
\vspace*{\floatsep}
\includegraphics[scale=0.45]{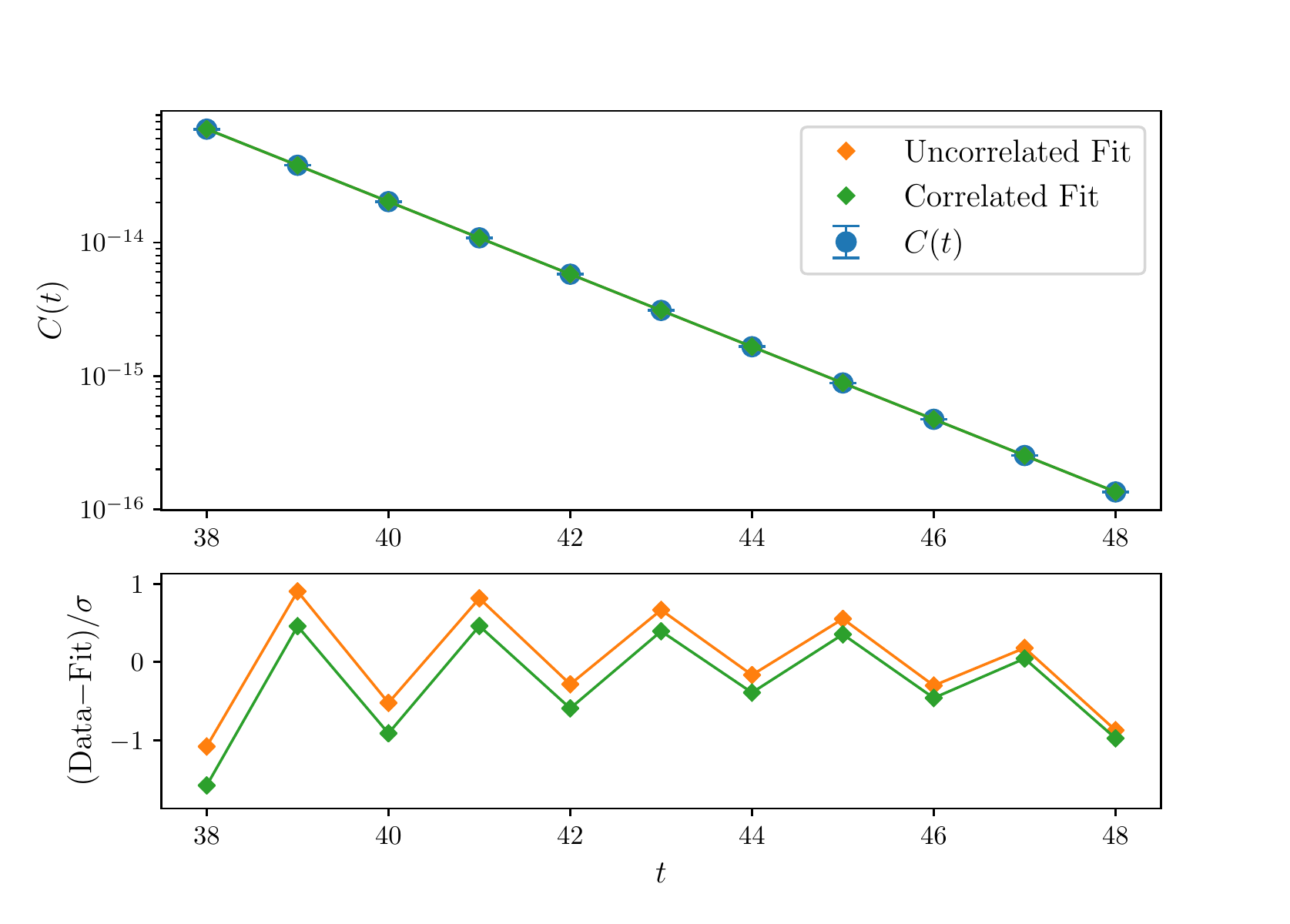}
\includegraphics[scale=0.45]{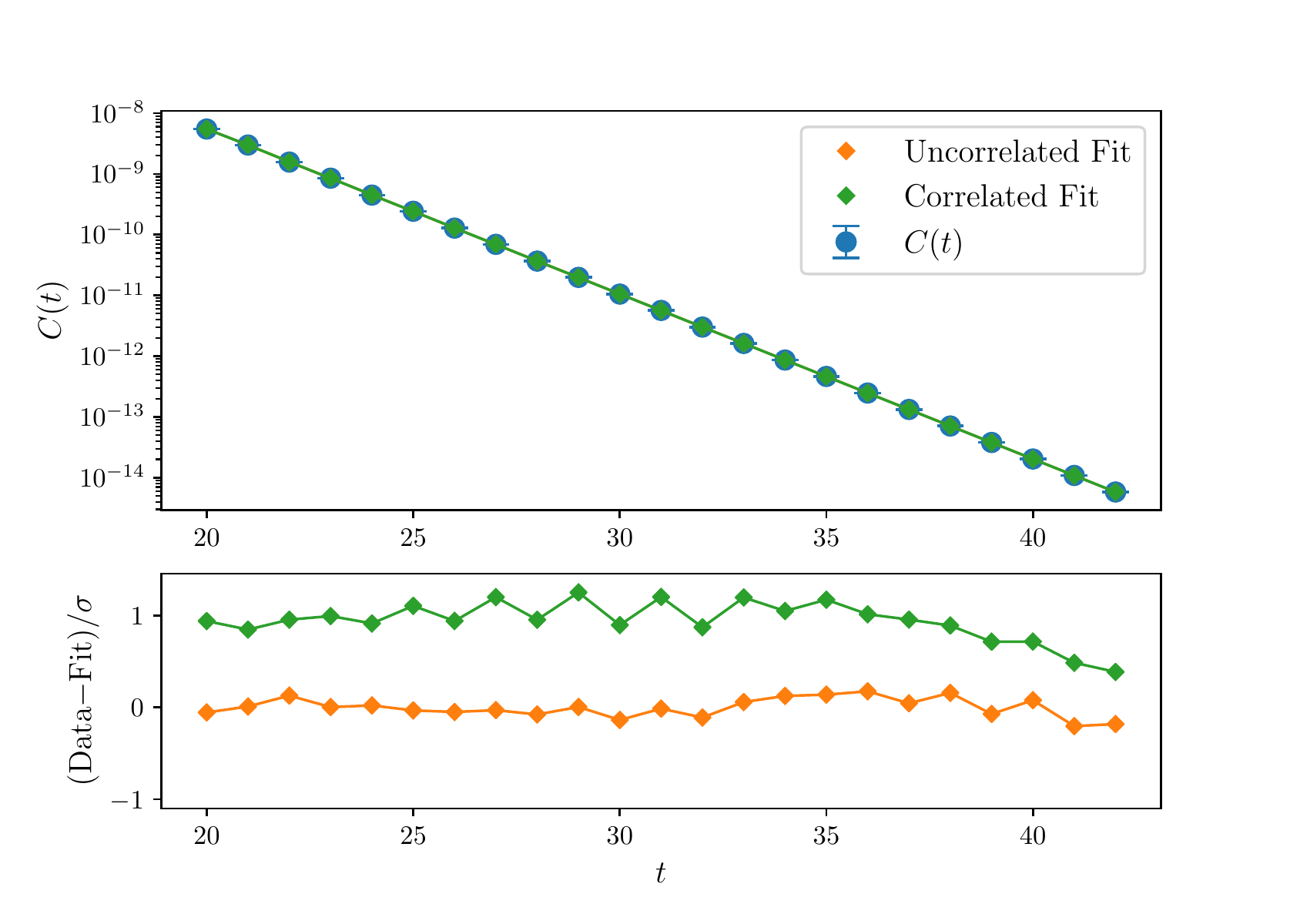}
\caption{Single-state fit without oscillation \texttt{exp}: $f(t)=c_0 (e^{-c_1 t}+ e^{-c_1 (T-t)})$ (left) and two-state fit including oscillations \texttt{expexpcos}: $f(t)=c_0 (e^{-c_1 t}+ e^{-c_1 (T-t)})+c_2 (e^{-c_3 t}+ e^{-c_3 (T-t)})+c_4 (e^{-c_5 t}+ e^{-c_5 (T-t)})\cos(\pi t)$ (right) of the correlator $C(t)$ in the \texttt{sc}-channel. The bias (data minus fit) is shown in the lower panels.\label{fig6}}
\end{figure}

\begin{figure}[h]
\center
\includegraphics[scale=0.474]{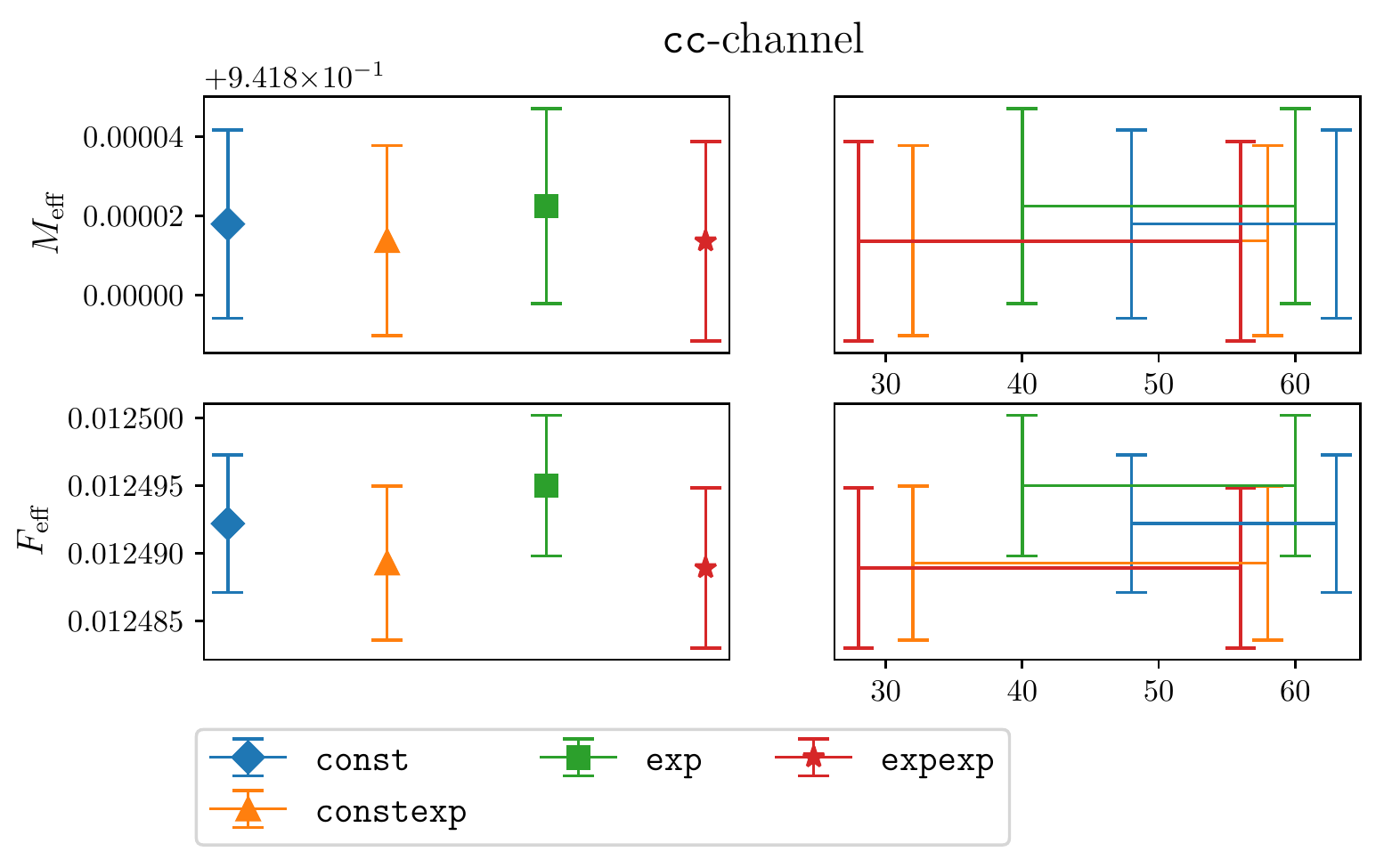}
\includegraphics[scale=0.474]{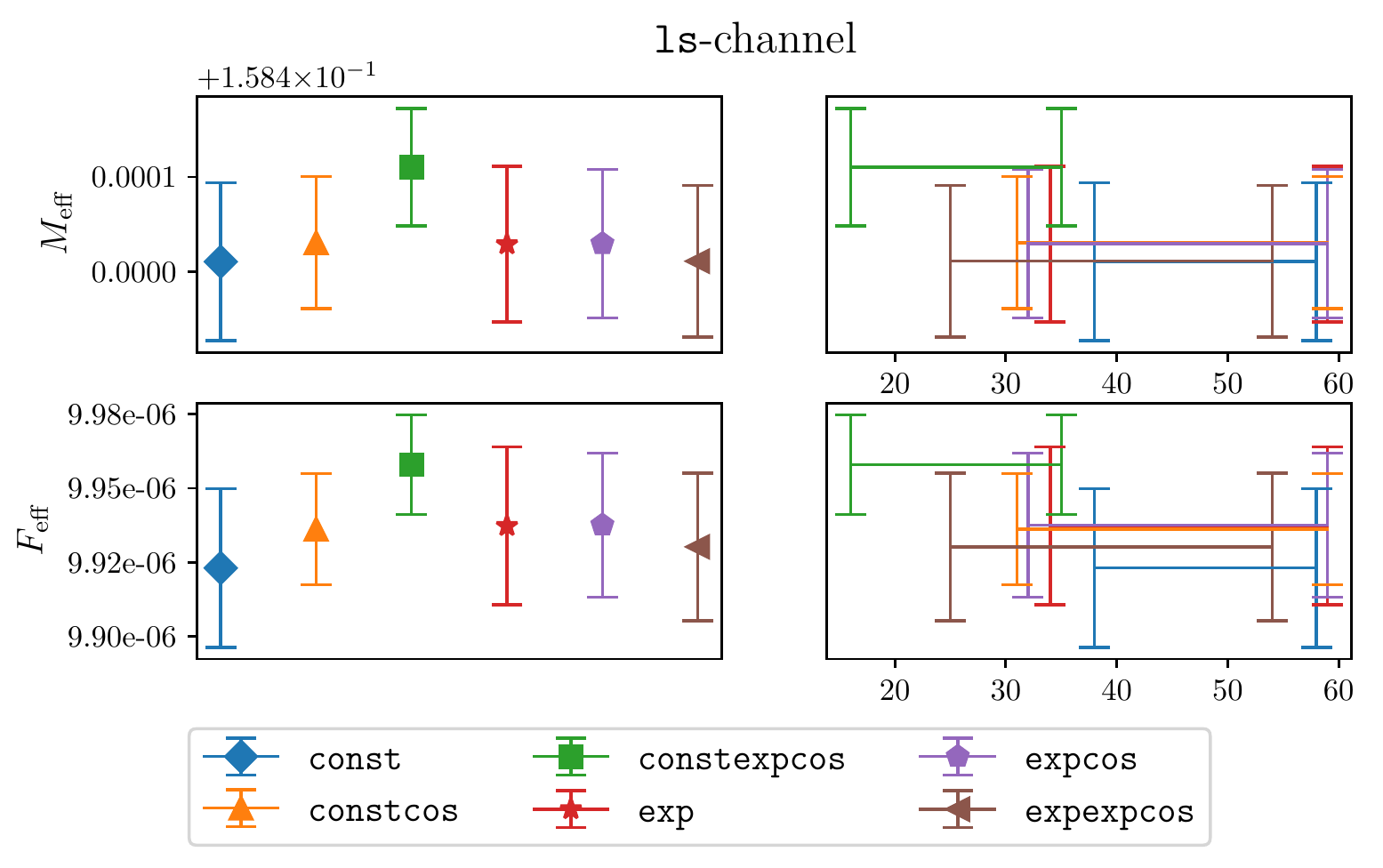}
\caption{Two examples of a summary plot, showing the values of the effective mass and effective decay constant obtained from various fits. The labels  \texttt{const, constexp, constcos, constexpcos} refer to fits of $\meff(t)$ and $\feff(t)$, while \texttt{exp, expexp, expcos, expexpcos} refer to one- or two-state fits of the Correlator $C(t)$. The panels on the right show the $t$-ranges over which the fits were performed.\label{fig7} }
\end{figure}

Having chosen a definition of $\meff(t)$ and $\feff(t)$ (we used the "acosh" definition for all channels) we first perform a constant fit inside the plateau region. In order to find a suitable fit range we plot the $Q$ value (see Figure \ref{fig3}) as an indicator of the quality of fit. We can then choose a suitable fit range from somewhere near the middle of the triangular region of large $Q$-values. Similarly one can plot the resulting fit parameters (i.e.\ $c_0$ in this case) to make sure there are no jumps inside this region. The $Q$-value is given by an incomplete Gamma-function $Q=\Gamma\left(\frac{m}{2},\frac{\chi^2}{2}\right)/\Gamma\left(\frac{m}{2}\right)$ with $m=(t_\mathrm{max}-t_\mathrm{min}+1)-n_\mathrm{par}$, where $n_\mathrm{par}$ is the number of parameters in the fit function.\\
After obtaining a first value for $\meff$ and $\feff$ from the constant fits, we expand our fit function to include an exponentially decreasing term to take into account the contributions from excited states (see Figure \ref{fig4}).  This extends our triangle of good fit ranges considerably to smaller $\tmin$ (see Figure \ref{fig3}). For the channels with oscillatory behaviour (\texttt{sc,lc,ls}) an exponentially damped cosine term should be included (see Figure \ref{fig5}).\\
With the values obtained from fitting $\meff(t)$ and $\feff(t)$ as initial values, the correlator itself can be fitted. First only the ground state (one cosh function) and then including the first and maybe the second excited state (two or three cosh functions). As before, in pseudo-scalar channels with unequal quark masses, a cosine term is required (see Figure \ref{fig6}).\\
Figure \ref{fig7} shows two examples of a summary plot  to compare the values of $\meff$ and $\feff$ obtained from the various fits to be used in the estimation of the systematic error.
\section{Synthetic data}
\begin{figure}
\includegraphics[scale=0.42]{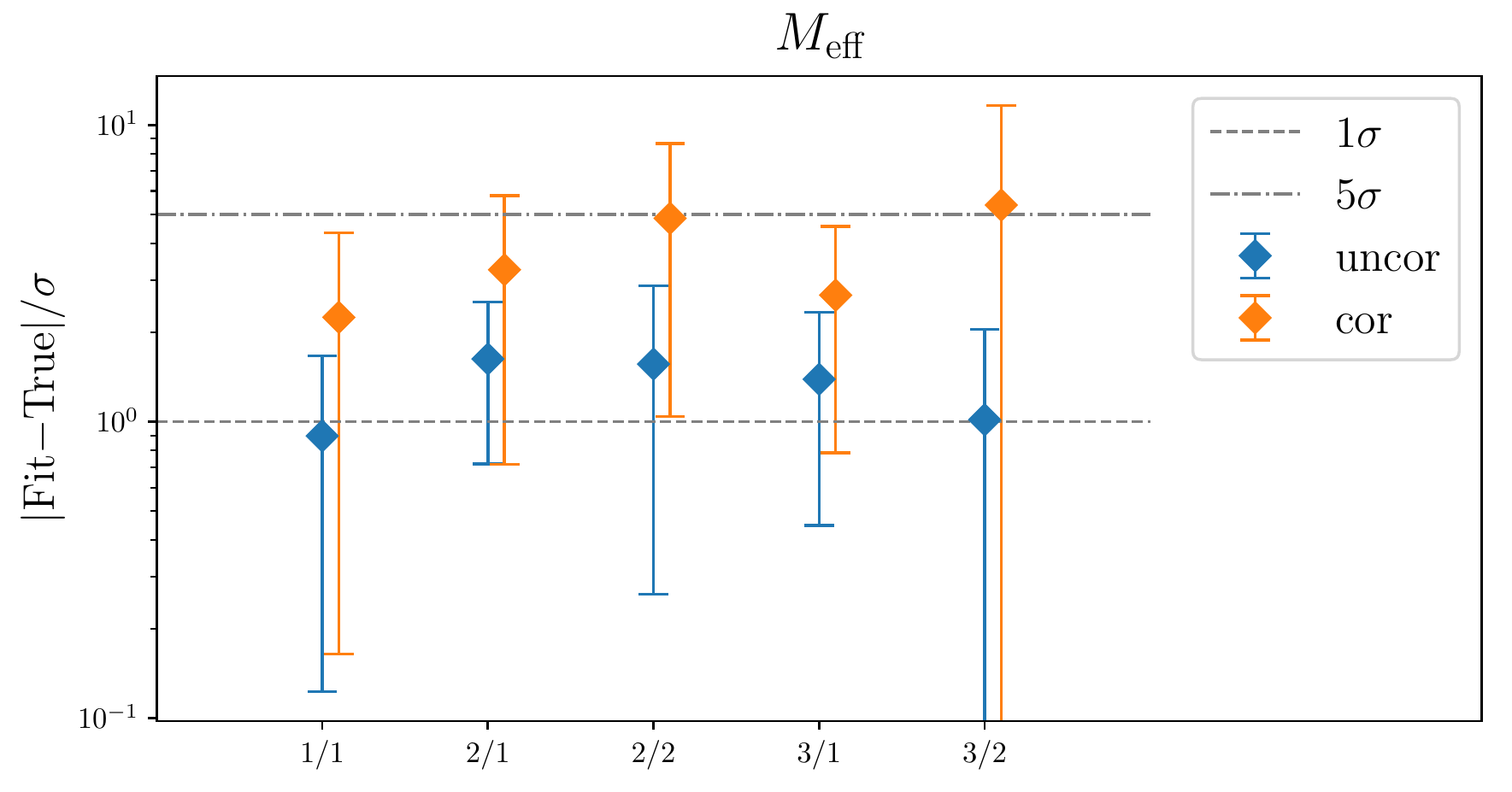}
\includegraphics[scale=0.42]{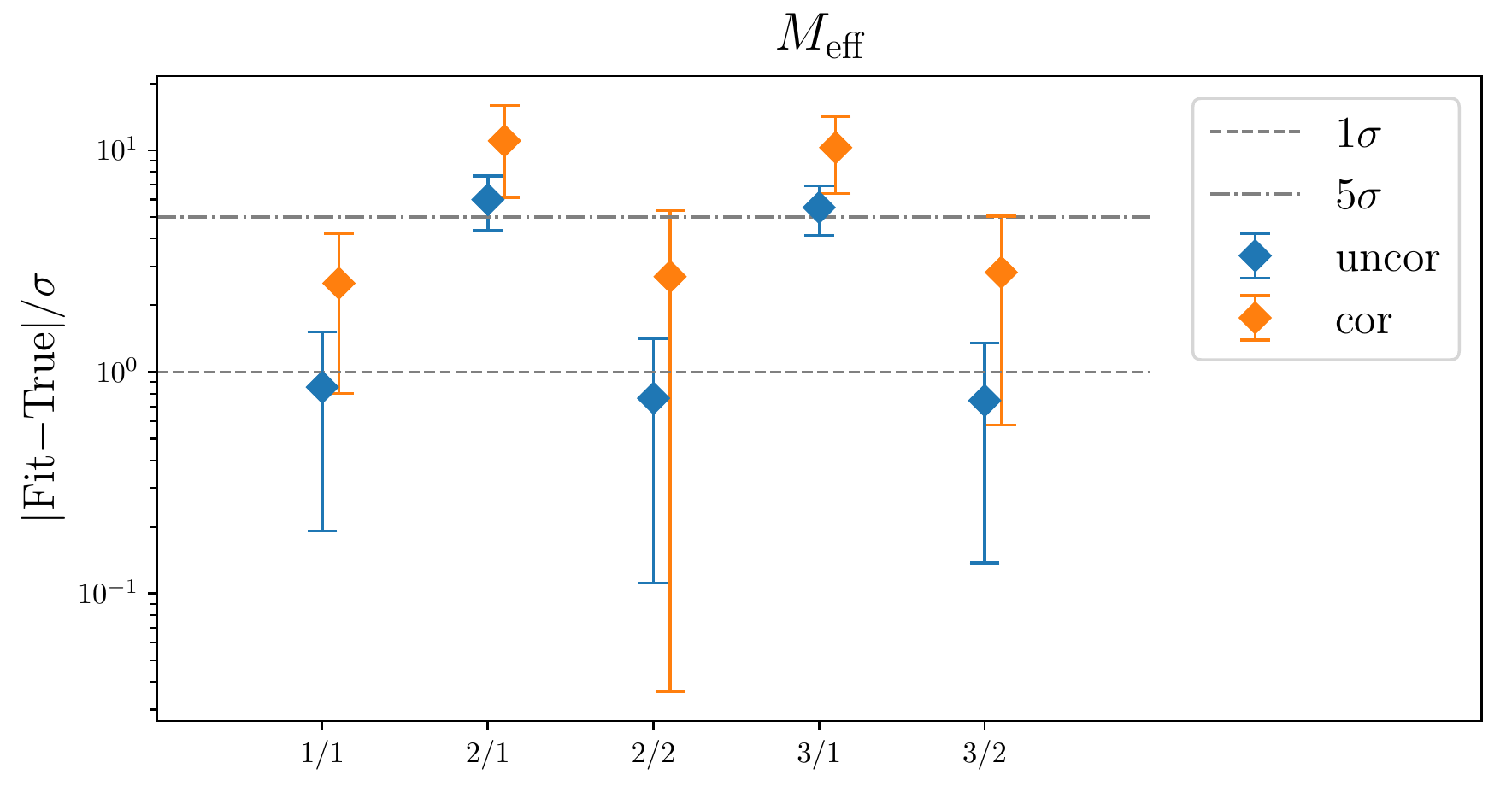}\\
\center
\includegraphics[scale=0.42]{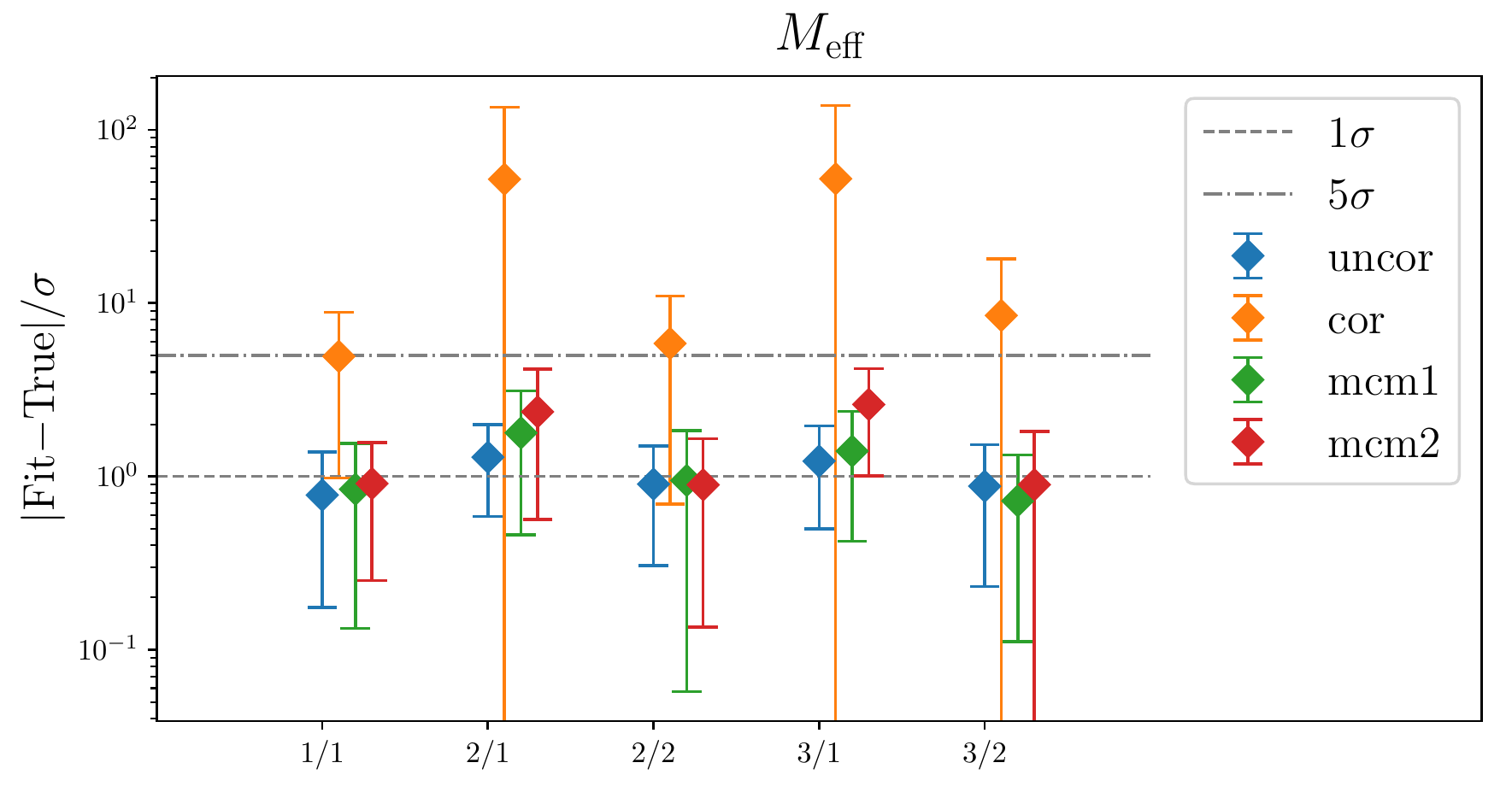}
\caption{Comparison of the biases of uncorrelated and correlated fits.  Shown are the absolute deviations of the fit results for $\meff$ from the true values in units of $\sigma$. Each data point corresponds to the mean of 50 sets of synthetic data. For the top left plot no correlation between different time slices has been generated. For  the top right plot the correlation is exponentially decreasing in $\Delta t$. For the bottom plot the correlation was generated by using the covariance matrices from actual simulated data (specifically the \texttt{cc}-channel from before). The labels \texttt{mcm1}  and \texttt{mcm2} stand for a method of simplifying the covariance matrix by only keeping the largest one or two of its singular values. The labels on the $x$-axis indicate the number of states in the data (left of the "/") and the number of states in the fit function (right of the "/").
\label{fig8}}
\end{figure}

In the plots above, both the results from correlated and uncorrelated fitting were shown. The correlated fit either agrees with the uncorrelated one or it may deviate from the data quite considerably. We have observed such large biases quite often in the correlated fits. They also exhibit erratic behaviour under small changes of the fit range $[\tmin,\tmax]$. These are known issues with correlated fitting \cite{Michael:1993yj,Jang:2011fp,Kilcup:1993ur,Seibert:1993sf}. If the number of independent configurations $n$, that go into the calculation of the sample covariance matrix, is not much larger than the length of the fitting interval $p=\tmax-\tmin$, it will underestimate the errors of the fit parameters and will lead to a very badly conditioned covariance matrix. The numerical inversion of such a badly conditioned matrix will lead to the biases in the correlated fits. In our case $p=10\sim 30$ and $n=31$, hence we cannot expect correlated fitting to work very well. To find out how accurate correlated fitting can be on data similar to ours, we generated synthetic correlator data with the same dimensions as in the analysis above. We used three different methods to generate noise on top of the correlator, whose exact mass is known. In Figure \ref{fig8} the biases of uncorrelated and correlated fits on the synthetic data sets are shown, depending on the number of states in the data and in the fitting functions. The correlated fits always perform worse than the uncorrelated ones, especially so when there are more states in the data than being fitted, which is the more realistic case. We have also used a number of methods that somehow truncate or modify the covariance matrix (see e.g.\ \cite{Bernard:2002pc} and references therein).
The results of two of them are also shown in Figure \ref{fig8}, but none of them would improve on the uncorrelated results. 
\section{Conclusion}
We compared a variety of fitting methods to extract effective masses and decay constants of pseudo-scalar mesons from staggered fermion data. In practice correlated fitting can introduce large biases and is less reliable than uncorrelated fitting. Using synthetic data, we showed that correlated fitting cannot be expected to give satisfactory results for data similar to ours. All statistical errors and covariances needed in a physics analysis can be obtained from the jackknife, even when using only the results of uncorrelated fits.
\section*{Acknowledgments}
\noindent
We thank the BMW-collaboration for allowing us to use their data; production details are found in Refs. \cite{Borsanyi:2016lpl,Borsanyi:2013bia}. M.A.\ thanks the DFG for funding under the collaborative scheme SFB-TRR 55.

\end{document}